\documentclass[11pt]{article}
\usepackage{epsfig} 
\newcommand{\pslash}{p \! \! \! /} 
 
\newcommand{\partialslash}{\partial \! \! \! /} 
\newcommand{\half}{\mbox{\small{$\frac{1}{2}$}}} 
\newcommand{\Nc}{N_{\!c}} 
\newcommand{\Nf}{N_{\!f}} 
\newcommand{\MSbar}{\overline{\mbox{MS}}} 

\begin{document}
\title{The Curci-Ferrari model with massive quarks at two loops} 
\author{R.E. Browne \& J.A. Gracey, \\ Theoretical Physics Division, \\ 
Department of Mathematical Sciences, \\ University of Liverpool, \\ Peach 
Street, \\ Liverpool, \\ L69 7ZF, \\ United Kingdom.} 
\date{} 
\maketitle 
\vspace{5cm} 
\noindent 
{\bf Abstract.} Massive quarks are included in the Curci-Ferrari model and the
theory is renormalized at two loops in the $\MSbar$ scheme in an arbitrary
covariant gauge.  

\vspace{-14cm} 
\hspace{13.5cm} 
{\bf LTH 542} 

\newpage 

Recently there has been renewed interest in examining covariantly gauge fixed
QCD where the gluon and ghost fields are given explicit mass terms. For 
instance, in \cite{1,2,3} such models have been used to investigate how a mass 
gap emerges for the gluon. In particular a dimension two operator obtains a 
non-zero vacuum expectation value which generates a gluon mass. Indeed in this 
context the effective potential calculation of \cite{3} demonstrated that the 
non-perturbative vacuum favoured a non-zero vacuum expectation value for the 
simplest dimension two operator possible in Yang-Mills theories. In the main 
these studies did not involve quarks and were effectively based on a version of
Yang-Mills theory with massive gluons originally introduced by Curci and 
Ferrari in \cite{4}. There a renormalizable theory of massive gluons was 
constructed with the aim of being an alternative to the Higgs mechanism for 
endowing vector bosons with mass. The main shortcoming of the Curci-Ferrari 
model, however, was the breaking of unitarity directly as a result of the 
massive gluon, \cite{5,6,7,8,9,10}. Nevertheless the model has proved useful 
for a variety of reasons. From a field theoretic point of view it is of 
interest due to the nonlinear nature of the gauge fixing term which introduces 
a quartic ghost self-interaction as well as modifying the usual ghost gluon 
interaction present in a linear covariant gauge fixing, \cite{4}. The nonlinear
property has been examined in \cite{10,11}. Moreover, there has been a debate 
on the BRST symmetry of the Curci-Ferrari model. For instance, Ojima, \cite{6},
has carried out a comprehensive examination of the BRST algebra with and 
without a mass term and explicitly constructed a negative norm state which 
therefore supports the lack of unitarity in the model. Other such states have 
been determined in \cite{8}. Another feature which emerged in these papers was 
the non-nilpotency of the BRST charge which appears to follow as a consequence 
of the exact form of the Lagrangian. A more recent study of this has been given
in \cite{12}. A final motivation for considering massive gluons rests in 
phenomenological considerations. For example, lattice results, \cite{13}, 
suggest that at low energies the gluon is massive though the precise form of 
the propagator at this scale is not known explicitly but is clearly dependent 
on non-perturbative properties. However, one can use a Curci-Ferrari type 
theory with the hope that it can provide useful insight into physics which is 
in someway dependent on a gluon which is massive. For instance, diffractive 
scattering has been examined in \cite{14} with such a motivation. 

Whilst the majority of the papers in this area concentrate on Yang-Mills theory
with a massive gluon the real world is based on QCD which involves quarks. 
Therefore, if complete studies of the property of a massive gluon are to be
performed, massive quarks need to be included. In \cite{6} matter fields were 
considered but were shown not to affect the failings of the Curci-Ferrari model
such as lack of unitarity or non-nilpotency of the BRST charge. The 
multiplicative renormalizabilty of the model, which was proved in 
\cite{4,8,9,10}, was unaffected. Given that the Curci-Ferrari model has been 
renormalized at one loop in \cite{8,12,15} and more recently at two loops in 
$\MSbar$ in \cite{16}, the purpose of this letter is to extend the 
renormalization of the Curci-Ferrari model at two loops to the case where 
massive quarks are included. This is important for various reasons. First, the 
main motivation of \cite{16} was to provide a calculational tool for 
renormalizing Green's functions of Yang-Mills theories where one could not 
naively nullify the momenta of several external legs without introducing 
spurious infrared infinities. Ordinarily one can handle such infrared problems
if some form of infrared rearrangement, \cite{15,16}, is performed to eliminate
those divergences which cannot be distinguished from ultraviolet ones in a 
dimensionally regularized calculation. Moreover, given that infrared 
rearrangement is usually carried out by hand it does not lend itself readily to
automatic calculations by computer. One simple resolution of the infrared 
problem is to introduce an infrared mass regularization. However, in using the 
Curci-Ferrari model, \cite{4}, which {\em naturally} incorporates a gluon and 
ghost mass and preserves renormalizability one resolves the infrared problem, 
\cite{19,10,15}, and opens the path to automatic calculations. Therefore, 
extending the Curci-Ferrari model to include quarks is the natural way to 
proceed in order to provide a tool similar to \cite{14} for QCD. However, there
is an additional practical point of view for our work. In conventional QCD 
perturbative calculations the fields of the Lagrangian are massless. However, 
in the real world the quarks have a physical mass. Therefore, including an 
explicit mass term for the quarks which is {\em independent} of the gluon mass 
we will have a Lagrangian which in principle includes as a special case the 
more realistic situation of massive quarks and {\em massless} gluons and ghosts
as well as allowing us to interpolate between various different scenarios. For 
the two loop $\MSbar$ renormalization of the Lagrangian we do not expect 
significant differences with conventional results for the renormalization group
functions. However, the tool we provide here will, for example, be useful in 
studying the mixing of operators under renormalization to operators of the same
and {\em lower} dimension since we will have an explicit quark mass at our 
disposal. 

We take as our Lagrangian, \cite{4,6},  
\begin{eqnarray} 
L &=& -~ \frac{1}{4} G_{\mu\nu}^a G^{a \, \mu\nu} ~-~ \frac{1}{2\alpha} 
(\partial^\mu A^a_\mu)^2 ~+~ \frac{m^2}{2} A_\mu^a A^{a \, \mu} ~+~ 
\partial_\mu \bar{c}^a \partial^\mu c^a ~-~ \alpha m^2 \bar{c}^a c^a 
\nonumber \\
&& -~ \frac{g}{2} f^{abc} A^a_\mu \, \bar{c}^b \, {\stackrel \leftrightarrow 
{\partial^\mu} } \, c^c ~+~ \frac{\alpha g^2}{8} f^{eab} f^{ecd} \bar{c}^a c^b 
\bar{c}^c c^d \nonumber \\ 
&& +~ i \bar{\psi}^{iI} \partialslash \psi^{iI} ~-~ \sqrt{\beta} m 
\bar{\psi}^{iI} \psi^{iI} ~-~ g \bar{\psi}^{iI} \gamma^\mu T^a_{IJ} \psi^{iJ} 
A^a_\mu 
\label{lag}
\end{eqnarray}  
where $1$ $\leq$ $a$ $\leq$ $N_A$, $1$ $\leq$ $I$ $\leq$ $N_F$ with $N_F$ and
$N_A$ the dimensions of the colour group fundamental and adjoint 
representations respectively where the structure constants are $f^{abc}$, $1$ 
$\leq$ $i$ $\leq$ $\Nf$ where $\Nf$ is the number of quark flavours, $g$ is the
coupling constant, $m$ is the gluon mass and hence the basic mass scale of the 
classical theory and $\bar{c}^a \, {\stackrel \leftrightarrow {\partial_\mu} } 
\, c^b$ $=$ $\bar{c}^a \partial_\mu c^b$ $-$ $(\partial_\mu \bar{c}^a) c^b$. 
The field strength, $G^a_{\mu\nu}$, follows from the definition of the 
covariant derivative as $G^a_{\mu\nu}$ $=$ $\partial_\mu A^a_\nu$ $-$ 
$\partial_\nu A^a_\mu$ $-$ $g f^{abc} A^b_\mu A^c_\nu$. We have included the 
usual covariant gauge fixing term with parameter $\alpha$ which, to ensure that
the action is BRST invariant, \cite{4,6,8,10}, dictates the form of the ghost 
interactions with a gluon of mass $m$ and a ghost of mass $\sqrt{\alpha} m$. To
make the two loop calculations easier to perform we have chosen to parameterize 
the quark mass with the parameter $\beta$ so that there is one basic mass 
parameter. This means that $\beta$ will get renormalized but the full quark 
mass, $\sqrt{\beta} m$, will be renormalized in such a way that it is 
independent of $\alpha$. The case of massive quarks but massless gluons is 
recovered by setting $m$ $\rightarrow$ $0$ in such a way that $\sqrt{\beta} m$ 
remains finite. With this Lagrangian the Minkowski space propagators for the 
gluon, ghost and quark fields are respectively,  
\begin{equation}
-~ i \delta^{ab} \left[ \frac{\eta^{\mu\nu}}{(k^2-m^2)} ~-~ 
\frac{(1-\alpha)k^\mu k^\nu}{(k^2-m^2)(k^2-\alpha m^2)} \right] ~~,~~ 
\frac{i\delta^{ab}}{(k^2-\alpha m^2)} ~~,~~ 
\frac{i\delta^{ij}(\pslash-\sqrt{\beta}m)}{(p^2-\beta m^2)} ~.  
\label{propdefn} 
\end{equation} 

The renormalization of (\ref{lag}) proceeds along usual grounds. First, we 
introduce the renormalized variables via 
\begin{eqnarray} 
A^{a \, \mu}_{\mbox{\footnotesize{o}}} &=& \sqrt{Z_A} \, A^{a \, \mu} ~~,~~ 
c^a_{\mbox{\footnotesize{o}}} ~=~ \sqrt{Z_c} \, c^a ~~,~~ 
\bar{c}^a_{\mbox{\footnotesize{o}}} ~=~ \sqrt{Z_c} \, \bar{c}^a ~~,~~ 
\psi_{\mbox{\footnotesize{o}}} ~=~ \sqrt{Z_\psi} \psi \nonumber \\ 
g_{\mbox{\footnotesize{o}}} &=& Z_g \, g ~~,~~ m_{\mbox{\footnotesize{o}}} ~=~ 
Z_m \, m ~~,~~ \alpha_{\mbox{\footnotesize{o}}} ~=~ Z^{-1}_\alpha Z_A \, 
\alpha ~~,~~ \beta_{\mbox{\footnotesize{o}}} ~=~ Z_\beta \beta 
\label{Zdefns}
\end{eqnarray} 
where the subscript, ${}_{\mbox{\footnotesize{o}}}$, denotes bare quantities.
Since our calculation builds on the Yang-Mills version of the Curci-Ferrari 
model we have not assumed the usual Slavnov-Taylor identity of 
$Z_\alpha$~$=$~$1$. Therefore, we have seven independent renormalization 
constants to compute. However, we do have various cross-checks on the results 
we will obtain for the renormalization constants. Given that the parameter $m$ 
appears in both the gluon and ghost sectors we ought to obtain a consistent 
renormalization for it from considering independently the gluon and ghost two 
point functions. Moreover, we can check that our wave function renormalizations
are correct by examining the various vertex corrections. The same coupling 
constant renormalization constant ought to emerge in all cases. Given the 
explicit values of the renormalization constants the corresponding 
renormalization group functions follow from the relations 
\begin{eqnarray} 
\gamma_A(a) &=& \beta(a) \frac{\partial \ln Z_A}{\partial a} ~+~
\alpha \gamma_\alpha(a) \frac{\partial \ln Z_A}{\partial \alpha} \nonumber \\
\gamma_\alpha(a) &=& \left[ \beta(a) \frac{\partial \ln Z_\alpha}{\partial a} 
{}~-~ \gamma_A(a) \right] \left[ 1 ~-~ \alpha \frac{\partial 
\ln Z_\alpha}{\partial \alpha} \right]^{-1} \nonumber \\ 
\gamma_m(a) &=& -~ \beta(a) \frac{\partial \ln Z_m}{\partial a} ~-~
\alpha \gamma_\alpha(a) \frac{\partial \ln Z_m}{\partial \alpha} \nonumber \\
\gamma_\beta(a) &=& -~ \beta(a) \frac{\partial \ln Z_\beta}{\partial a} ~-~
\alpha \gamma_\alpha(a) \frac{\partial \ln Z_\beta}{\partial \alpha} 
\label{rgedefn} 
\end{eqnarray} 
where $\gamma_\beta(a)$ is defined by
\begin{equation}
\gamma_\beta(a) ~=~ \frac{\mu}{\beta} \frac{\partial \beta}{\partial \mu} 
\end{equation} 
and $a$ $=$ $g^2/(16\pi^2)$. (The parameter $\beta$ ought not to be confused 
with the $\beta$-function, $\beta(a)$.) In ordinary QCD one would have 
$\gamma_A(a)$ $=$ $-$ $\gamma_\alpha(a)$ but since $Z_\alpha$ $\neq$ $1$ in the
Curci-Ferrari model we do not expect this to be restored in the presence of 
quarks. Moreover, this is the origin of the second terms in the expressions for
$\gamma_m(a)$ and $\gamma_\beta(a)$ which will be dependent on $\alpha$.  

Given that we are renormalizing a massive version of QCD we need to have an
algorithm for computing the ultraviolet divergences of the massive multiscale
two loop Feynman integrals which contribute. First, we use dimensional 
regularization with $d$ $=$ $4$ $-$ $2\epsilon$. Second, we follow the strategy
of \cite{20,21} where massive Feynman integrals are expanded in powers of the
external momenta based on the identity
\begin{equation} 
\frac{1}{((k-p)^2-m^2)} ~=~ \frac{1}{(k^2-m^2)} ~+~ \frac{(2kp - p^2)}
{((k-p)^2-m^2)(k^2-m^2)} ~.  
\end{equation} 
The expansion is terminated by the rule that when the powers of momenta exceed 
those which can appear in the Green's function through renormalizability, then 
they are dropped. So, for example, the gluon propagator is only expanded to 
$O(p^2)$ with the $O(p^3)$ terms being dropped where $p$ is the external 
momentum. Consequently, one is left with massive vacuum bubble graphs where 
because of the different masses of (\ref{lag}), not all the masses are equal. 
However, massive vacuum two loop bubbles have been evaluated to the finite part
in \cite{22}, for example, though we only require the Laurent expansion to the 
simple pole in $\epsilon$. For instance, the basic vacuum bubble with three 
different scales is given by  
\begin{eqnarray} 
\int_{kl} \frac{1}{(k^2-m^2)(l^2-\alpha m^2)[(k-l)^2-\beta m^2]} &=& 
\left[ ~-~ (1 + \alpha + \beta) \left( \frac{1}{2\epsilon^2} ~+~ 
\frac{3}{2\epsilon} ~+~ \frac{1}{\epsilon} \ln \left( \frac{4\pi}{m^2 e^\gamma}
\right) \right) \right. \nonumber \\
&& \left. ~~+~ \left( \alpha \ln \alpha ~+~ \beta \ln \beta \right) 
\frac{1}{\epsilon} ~+~ O(1) \right] \frac{m^2}{(4\pi)^4} 
\end{eqnarray}
where $\gamma$ is the Euler-Mascheroni constant and $\int_k$ $=$ $\int 
d^d k/(2\pi)^d$. The expression for different powers of the propagators are 
determined by differentiating with respect to the parameter $\alpha$, $\beta$
and $m^2$. It is worth noting that the appearance of $\ln \alpha$ and $\ln 
\beta$ terms could in principle lead to a non-analytic renormalization 
constant. However, these ought to cancel when all contributions from one and
two loop graphs are included. In addition, in reducing the integrals to vacuum
bubbles, partial fractions have been used which can give rise to other 
potentially singular terms. For instance, 
\begin{equation} 
\frac{1}{(k^2-\alpha m^2)(k^2-\beta m^2)} ~=~ \frac{1}{(\alpha-\beta)m^2} 
\left[ \frac{1}{(k^2-\alpha m^2)} ~-~ \frac{1}{(k^2- \beta m^2)} \right] ~.  
\end{equation} 
This provides another internal check since there are no singular terms in the
original Lagrangian and for (\ref{lag}) to be renormalizable they ought not to
remain after the two loop calculation. Given the large amount of algebra which
arises due to the expansion to vacuum bubbles, to handle their evaluation for 
different masses we have written an algorithm in a symbolic manipulation 
language, {\sc Form} version $3$, \cite{23}. The calculation proceeds 
automatically, since, for example the Feynman diagrams are generated using the 
{\sc Qgraf} package, \cite{24}. Moreover, the renormalization constants are 
extracted by computing the Green's functions in terms of bare parameters then 
rescaling by (\ref{Zdefns}) after the pole structure has been determined and 
following the procedure of \cite{25}. This reproduces the usual method of 
subtractions automatically. 

We have renormalized the gluon, ghost and quark two-point functions as well
as the three $3$-point vertices. We obtain the following renormalization 
constants 
\begin{eqnarray} 
Z_A &=& 1 ~+~ \left[ \left( \frac{13}{6} - \frac{\alpha}{2} \right) C_A 
- \frac{4}{3} T_F \Nf \right] \frac{a}{\epsilon} \nonumber \\
&& +~ \left[ \left( \left( \frac{3\alpha^2}{16} - \frac{17\alpha}{24} 
- \frac{13}{8} \right) C_A^2 + C_A T_F\Nf \left( \frac{2}{3}\alpha + 1 \right) 
\right) \frac{1}{\epsilon^2} \right. \nonumber \\
&& \left. ~~~~~-~ \left( \left( \frac{\alpha^2}{16} + \frac{11\alpha}{16}
- \frac{59}{16} \right) C_A^2 + 2 C_F T_F\Nf + \frac{5}{2} C_A T_F \Nf \right) 
\frac{1}{\epsilon} \right] a^2 ~+~ O(a^3) \nonumber \\ 
Z_\alpha &=& 1 ~-~ \left( \frac{\alpha}{4} \right) C_A \frac{a}{\epsilon} ~+~ 
C_A^2 \left[ \left( \frac{\alpha^2}{16} + \frac{3\alpha}{16} 
\right) \frac{1}{\epsilon^2} ~-~ \left( \frac{\alpha^2}{32} 
+ \frac{5\alpha}{32} \right) \frac{1}{\epsilon} \right] a^2 ~+~ O(a^3) 
\nonumber \\ 
Z_c &=& 1 ~+~ \left( \frac{3}{4} - \frac{\alpha}{4} \right) C_A 
\frac{a}{\epsilon} ~+~ \left[ \left( \left( \frac{\alpha^2}{16} - \frac{35}{32}
\right) C_A^2 + \frac{1}{2} C_A T_F \Nf \right) \frac{1}{\epsilon^2} \right. 
\nonumber \\ 
&& \left. -~ \left( \left( \frac{\alpha^2}{32} - \frac{\alpha}{32} 
- \frac{95}{96} \right) C_A^2 + \frac{5}{12} C_A T_F \Nf \right) 
\frac{1}{\epsilon} \right] a^2 ~+~ O(a^3) \nonumber \\ 
Z_\psi &=& 1 ~-~ \alpha C_F \frac{a}{\epsilon} ~+~ \left[ \left( C_F C_A  
\left( \frac{\alpha^2}{8} + \frac{3\alpha}{4} \right) + \frac{\alpha^2}{2} 
C_F^2 \right) \frac{1}{\epsilon^2} \right. \nonumber \\ 
&& \left. -~ \left( C_F C_A \left( \alpha + \frac{25}{8} \right) ~-~ 
C_F T_F \Nf ~-~ \frac{3}{4} C_F^2 \right) \frac{1}{\epsilon} \right] a^2 ~+~ 
O(a^3) \nonumber \\ 
Z_m &=& 1 ~+~ \left[ \left( \frac{\alpha}{8} - \frac{35}{24} \right) C_A 
+ \frac{2}{3} T_F \Nf \right] \frac{a}{\epsilon} \nonumber \\
&& +~ \left[ \left( \left( - \frac{\alpha^2}{128} - \frac{53\alpha}{192} 
+ \frac{1435}{384} \right) C_A^2 ~+~ \frac{2}{3} T_F^2 \Nf^2 ~+~ \left( 
\frac{\alpha}{12} - \frac{19}{6} \right) C_A T_F \Nf \right) 
\frac{1}{\epsilon^2} \right.  \nonumber \\
&& \left. ~~~~~+~ \left( \left( \frac{\alpha^2}{64} + \frac{11\alpha}{64} 
- \frac{449}{192} \right) C_A^2 ~+~ C_F T_F \Nf ~+~ \frac{35}{24} C_A T_F \Nf 
\right) \frac{1}{\epsilon} \right] a^2 ~+~ O(a^3) \nonumber \\ 
Z_\beta &=& 1 ~+~ \left[ \left( \frac{35}{12} - \frac{1}{\alpha} \right) C_A  
- \frac{4}{3} T_F \Nf - 6 C_F \right] \frac{a}{\epsilon} \nonumber \\
&& +~ \left[ \left( \left( \frac{\alpha^2}{16} - \frac{13\alpha}{24} 
- \frac{35}{32} \right) C_A^2 ~+~ 18 C_F^2 ~+~ \left( \frac{\alpha}{3} 
+ \frac{1}{2} \right) C_A T_F \Nf \right. \right. \nonumber \\ 
&& \left. \left. ~~~~~~+~ 4 C_F T_F \Nf ~+~ \left( \frac{3\alpha}{2}
- \frac{13}{2} \right) C_F C_A \right) \frac{1}{\epsilon^2} ~-~ \left( 
\left( \frac{\alpha^2}{32} + \frac{11\alpha}{32} - \frac{449}{96} \right) C_A^2
\right. \right. \nonumber \\
&& \left. \left. ~~~~~~+~ \frac{3}{2} C_F^2 ~-~ \frac{4}{3} C_F T_F \Nf ~+~ 
\frac{35}{12} C_A T_F \Nf ~+~ \frac{97}{6} C_F C_A \right) \frac{1}{\epsilon} 
\right] a^2 ~+~ O(a^3) \nonumber \\ 
Z_g &=& 1 ~+~ \left( \frac{2}{3} T_F \Nf - \frac{11}{6} C_A \right) 
\frac{a}{\epsilon} ~+~ \left[ \left( \frac{121}{24}C_A^2 + \frac{2}{3} T_F^2
\Nf^2 - \frac{11}{3} C_A T_F \Nf \right) \frac{1}{\epsilon^2} \right. 
\nonumber \\ 
&& \left. +~ \left( C_F T_F \Nf + \frac{5}{3} C_A T_F \Nf - \frac{17}{6} 
C_A^2 \right) \frac{1}{\epsilon} \right] a^2 ~+~ O(a^3) 
\end{eqnarray}  
where $T^a T^a$ $=$ $C_F$, $f^{acd} f^{bcd}$ $=$ $C_A \delta^{ab}$ and 
$\mbox{Tr}(T^a T^b)$ $=$ $T_F \delta^{ab}$. We have recorded these explicitly
since the double pole in $\epsilon$ follows from the form of the one loop
simple pole and therefore provides another check on our computation. Moreover,
they are $\beta$-independent since we use a mass independent renormalization
scheme. From these values we obtain the renormalization group functions at two 
loops in $\MSbar$,  
\begin{eqnarray} 
\gamma_A(a) &=& \left[ ( 3\alpha - 13 ) C_A + 8T_F \Nf \right] \frac{a}{6} 
\nonumber \\
&& +~ \left[ \left( \alpha^2 + 11\alpha - 59 \right) C_A^2 + 40 C_A T_F \Nf 
+ 32 C_F T_F \Nf \right] \frac{a^2}{8} ~+~ O(a^3) \nonumber \\  
\gamma_\alpha(a) &=& -~ \left[ ( 3\alpha - 26 ) C_A + 16 T_F \Nf \right]
\frac{a}{12} \nonumber \\
&& -~ \left[ \left( \alpha^2 + 17\alpha - 118 \right) C_A^2 + 80 C_A T_F \Nf 
+ 64 C_F T_F \Nf \right] \frac{a^2}{16} ~+~ O(a^3) \nonumber \\ 
\gamma_c(a) &=& ( \alpha - 3 ) C_A \frac{a}{4} ~+~ \left[ \left( 3\alpha^2 
- 3\alpha - 95 \right) C_A^2 + 40 C_A T_F \Nf \right] \frac{a^2}{48} ~+~ O(a^3) 
\nonumber \\  
\gamma_\psi(a) &=& \alpha C_F a ~+~ C_F \left[ (8\alpha + 25)C_A - 6 C_F 
- 8 T_F \Nf \right] \frac{a^2}{4} ~+~ O(a^3) \nonumber \\  
\gamma_m(a) &=& \left[ ( 3\alpha - 35 ) C_A + 16 T_F \Nf \right] \frac{a}{24} 
\nonumber \\
&& +~ \left[ \left( 3\alpha^2 + 33\alpha - 449 \right) C_A^2 + 280 C_A T_F \Nf 
+ 192 C_F T_F \Nf \right] \frac{a^2}{96} ~+~ O(a^3) \nonumber \\
\gamma_\beta(a) &=& -~ \left[ ( 3\alpha - 35 ) C_A + 72 C_F + 16 T_F \Nf 
\right] \frac{a}{12} \nonumber \\
&& -~ \left[ \left( 3\alpha^2 + 33\alpha - 449 \right) C_A^2 + 1552 C_F C_A 
+ 280 C_A T_F \Nf \right. \nonumber \\
&& \left. ~~~~+~ 144 C_F^2 - 128 C_F T_F \Nf \right] \frac{a^2}{48} ~+~ 
O(a^3) \nonumber \\
\beta(a) &=& -~ \left[ \frac{11}{3} C_A - \frac{4}{3} T_F \Nf \right] a^2 ~-~ 
\left[ \frac{34}{3} C_A^2 - 4 C_F T_F \Nf - \frac{20}{3} C_A T_F \Nf \right]
a^3 ~+~ O(a^4) ~. \nonumber \\  
\end{eqnarray} 
The expression for the $\beta$-function agrees with the scheme independent 
results of \cite{26,27}. The renormalization group functions for the wave 
functions of the fields and $\alpha$ agree with the corresponding results of 
\cite{25,26,27,28} in the Landau gauge, $\alpha$ $=$ $0$. Indeed  
\begin{equation} 
\gamma_A(a) ~+~ \gamma_\alpha(a) ~=~ \alpha \left[ \frac{a}{4} C_A ~+~ 
\left( \alpha + 5 \right) C_A^2 \frac{a^2}{16} \right] ~+~ O(a^3) ~.  
\label{gamgam} 
\end{equation} 
which is independent of $\Nf$ and vanishes when $\alpha$ $=$ $0$. Moreover, 
(\ref{gamgam}) is equivalent to the statement that the ghost gluon vertex does
not get renormalized in the Landau gauge, \cite{29,30}, which was originally 
verified at one loop for (\ref{lag}) in \cite{29}. With the presence of quarks 
the gluon mass dimension now depends on $\Nf$ at one loop. However, to check we
have correctly determined the quark mass anomalous dimension we need to compute
the anomalous dimension of $m_q$~$=$~$\sqrt{\beta} m$. From  
\begin{equation}  
\gamma_{m_q}(a) ~=~ \gamma_m(a) ~+~ \half \gamma_\beta(a) 
\end{equation}  
we have 
\begin{equation} 
\gamma_{m_q}(a) ~=~ -~ 3 C_F a ~-~ C_F \left[ 97 C_A + 9 C_F - 20 T_F \Nf 
\right] \frac{a^2}{6} ~+~ O(a^3) 
\end{equation} 
which agrees with the two loop $\MSbar$ result of \cite{31,32} in our 
conventions and is independent of $\alpha$. For completeness, the ghost mass 
dimension is determined from 
\begin{equation} 
m^2_c ~=~ \alpha m^2 
\end{equation} 
which implies 
\begin{equation} 
\gamma_{m_c}(a) ~=~ \gamma_m(a) ~+~ \half \gamma_\alpha(a) 
\end{equation}  
giving 
\begin{equation} 
\gamma_{m_c}(a) ~=~ -~ \frac{3}{8} C_A a ~-~ \left[ \left( 18 \alpha + 95
\right) C_A^2 - 40 T_F \Nf \right] \frac{a^2}{96} ~+~ O(a^3) 
\end{equation} 
which becomes $\Nf$ and gauge dependent at two loops. Further checks on the
correctness of $\gamma_m(a)$ are that the Yang-Mills sector of our expression 
agrees with the three loop result of \cite{3} and the one loop expression of
\cite{33} where the multiplicative renormalizability of the composite operator 
$\half A^a_\mu A^{a \, \mu}$ $-$ $\alpha \bar{c}^a c^a$ was verified by 
determining the mixing matrix of the renormalization of the constituent 
dimension two operators. For completeness, we have evaluated $\gamma(a)$ in the 
Landau gauge for QCD, $\Nc$ $=$ $3$ and $T_F$ $=$ $\half$, and found
\begin{equation}
\left. \gamma_m(a) \right|_{\alpha=0} ~=~ \left( 8 \Nf - 105 \right) 
\frac{a}{24} ~+~ \left( 548 \Nf - 4041 \right) \frac{a^2}{96} ~+~ O(a^3) ~.  
\end{equation} 
It is worth noting that both coefficients are negative for $\Nf$ $<$ $8$ which
implies that the gluon mass runs to zero in the ultraviolet limit in this case. 

In conclusion we have provided the full two loop renormalization of the 
Curci-Ferrari model with massive quarks in the $\MSbar$ scheme. Whilst the 
nature of the model may appear unphysical with a gluon mass and lack of 
unitarity, it is important to recall that one long term aim is to use the the 
model to attack the renormalization of other Green's functions where quark mass
effects will become important and a {\em natural} infrared mass regularization 
is necessary. Further, it is worth noting that models with non-abelian 
symmetries and massive gluons may have important phenomenological consequences.
For instance, it has been shown in \cite{33} that including $O(1/Q^2)$ power 
corrections in the operator product expansion improves the matching of 
calculated physical quantities with experiment. Although such corrections 
derive from a gluon mass, albeit tachyonic in origin, they indicate the 
potential importance of massive gluons, such as that of (\ref{lag}), to probe 
non-perturbative physics.  


\end{document}